\documentclass[onecolumn, showpacs,amsmath,amssymb]{revtex4}
\usepackage{graphics}
\usepackage{amsmath}
\usepackage{amsfonts}
\usepackage{graphicx}
\usepackage{color}
\usepackage{epsfig}
\usepackage{epstopdf}
\usepackage{bm}

\begin{document}
\title{Cooperative excitation and many-body interactions in a cold Rydberg gas}
\author{Matthieu Viteau$^{1,*}$, Paul Huillery$^{1,2,3,*}$, Mark G. Bason$^{1,*}$,  Nicola Malossi$^{1,4}$, Donatella Ciampini$^{1,3,4}$,  Oliver Morsch$^1$, Ennio Arimondo$^{1,3,4}$,
        	     Daniel Comparat$^2$, \& Pierre Pillet$^2$}
\affiliation{$^1$INO-CNR, Largo Pontecorvo 3, 56127 Pisa, Italy\\  $^2$ Laboratoire Aim\'{e} Cotton, Universit\'e Paris-Sud 11, Campus d'Orsay Bat. 505, 91405 Orsay, France\\ $^3$Dipartimento di Fisica `E. Fermi', Universit\`a di
Pisa, Largo Pontecorvo 3, 56127 Pisa, Italy\\ $^4$CNISM UdR, Dipartimento di Fisica `E. Fermi', Universit\`a di Pisa, Largo Pontecorvo 3, 56127 Pisa, Italy
}

\begin{abstract}
The dipole blockade of Rydberg excitations is a hallmark of the strong interactions between atoms in these high-lying quantum states~\cite{Saffman2010,Comparat2010}. One of the consequences of the dipole blockade is the suppression of fluctuations in the counting statistics of Rydberg excitations, of which some evidence has been found in previous experiments. Here we present experimental results on the dynamics and the counting statistics of Rydberg excitations of ultra-cold Rubidium atoms both on and off resonance, which exhibit sub- and super-Poissonian counting statistics, respectively. We compare our results with numerical simulations using a novel theoretical model based on Dicke states of Rydberg atoms including dipole-dipole interactions, finding good agreement between experiment and theory.
\end{abstract}

\maketitle
Atoms excited to high-lying quantum states, so-called Rydberg atoms, are highly polarizable and, therefore, can interact strongly with each other at large distances~\cite{Saffman2010,Comparat2010}. The study of ultracold gases has opened up new avenues of the investigation of strong atomic interactions. A wealth of potential applications is associated with the manipulation of ultracold atoms excited to Rydberg states interacting through strong van der Waals (vdW) or long-range dipolar interactions, ranging from studies of strongly correlated quantum systems to quantum information~\cite{LesanovskyViewpoint2011}. A key signature of interactions between Rydberg atoms is the suppression of fluctuations in the number of excitations due to the dipole blockade.  While evidence of the dipole blockade has been observed in several experiments~\cite{CubelLiebisch2005,Reinhard2008,Gaetan2009,Urban2009}, prior studies were unable to demonstrate sub-Poissonian behaviour with a statistically significant confidence level. Also, those studies were carried out for resonant excitation only, leaving open the question of how Rydberg excitations in the strongly interacting regime evolve for finite detuning from resonance. \\
\indent The dipole blockade is a hallmark of the strong interactions between Rydberg atoms in an ultra-cold atomic gas. When an ensemble of atoms is irradiated by laser light resonant with a Rydberg excitation (either using a single laser or a multi-step excitation scheme), due
to the interaction between Rydberg atoms, the excitation of a particular atom can be suppressed by a neighbouring one that is already in a Rydberg state. The radius of influence of an atom in this sense is called the blockade radius, which depends on the interaction strength and the linedwidth of the Rydberg excitation. As a result of the dipole blockade the excitation dynamics of the atoms are strongly correlated, leading to a cut-off in the excitation when all the available blockade volumes in the sample have been exhausted \cite{Hernandez2008}. Another way of describing the phenomenon is to view the atoms within a blockade volume as a `superatom', meaning that rather than individual atoms, collective states of several atoms are excited \cite{Heidemannn07}. In the regime where the size of the sample is larger than the blockade radius, quantum correlations are expected to interconnect the whole sample. The description of this fully correlated regime is complex and the role of the atomic correlations on the atomic observables has recently received much attention~\cite{Comparat2010}. In a simple mean field-type model for the interacting Rydberg atoms \cite{Tong2004} each Rydberg atom experiences only an average interaction energy, which corresponds to neglecting all quantum correlations in the system. In order to reproduce the observed fluctuations and spatial correlations of the Rydberg excitations, however, more sophisticated approaches are needed ~\cite{Ates2006,Hernandez2006,Wuester2010,Stanojevic2010,Lesanovsky2011}.

\indent While the full dynamics of Rydberg excitations in a cold gas of thousands of atoms is difficult to access experimentally, a number of experiments have observed signatures of the dipole blockade \cite{Heidemannn07}, such as the suppression of the excitation of a single atom in the presence of a nearby Rydberg atom \cite{Gaetan2009,Urban2009}, the spatial correlations between Rydberg excitations \cite{Schwarzkopf2011} and indications of a suppression of fluctuations in the counting statistics near a F\"{o}rster resonance \cite{Reinhard2008}. In this Letter we present a comprehensive experimental and theoretical study of the counting statistics of Rydberg excitations of ultra-cold rubidium atoms in a magneto-optical trap (MOT).\\
\indent  As a full quantum mechanical treatment containing the states of all the  atoms in the MOT is not feasible,  the results of our experiments are analysed using an original theoretical model based on the well established Dicke model of quantum optics.  The Dicke model, originally introduced for describing cooperative spontaneous emission or superradiance~\cite{Dicke1954} and subradiance~\cite{Pavolini1985}, is here modified by including the vdW interactions between the Dicke collective states (DCS). Our approach  leads to a manageable size of the basis set for the simulations, which is only on the order of the number of Rydberg excitations expected, that is, around $10-20$ for our parameters, to be compared to our experimental numbers of ultracold atoms between tens and hundreds of thousands.   The collective Dicke states contain the full statistical information about the collective Rydberg excitation, not only the average quantities, and therefore allow calculation of all the moments of the excitation statistics, in particular the measured variance of the Rydberg excitations. Furthermore, including a coupling between Dicke states of different symmetry leads to a partially non-Markovian dynamics, which results in better agreement with experimental findings.  \\
\indent In our experiments, rubidium atoms in a MOT are excited to Rydberg states with principal quantum number $n$ between $50$ and $80$ using a two-photon scheme and detected, after field ionization, by a channeltron charge multiplier~\cite{Viteau2010} (for details see Supplemental Material). The detuning of the resulting coherent excitation from the ground state to a highly excited Rydberg state can be varied by changing the frequency of the second step laser at around $1015\,\mathrm{nm}$(the first step laser at $421\,\mathrm{nm}$ is detuned by around a GHz from an intermediate resonance in order to avoid population of that state). In the limit of vanishing interactions between the Rydberg atoms each excitation to a Rydberg state is independent of all the others, leading to a Poissonian excitation process. When the interactions are strong, however, the excitation processes are highly correlated and one expects sub-Poissonian counting statistics. These two regimes can be quantified through the Mandel Q-parameter~\cite{Mandel1979} defined as
\begin{equation}
Q=\frac{\langle\left(\Delta N\right)^2\rangle}{<N>}-1,
 \end{equation}
where $\langle \left(\Delta N\right)^2\rangle$ and $\langle N\rangle$ are the variance and mean, respectively, for the number $N$ of Rydberg excitations. Experimentally, we determine $Q$ by repeating an excitation-detection cycle for a fixed set of parameters between $50$ and $100$ times and extract the mean and the variance from that counting distribution. Clearly, for Poissonian processes $Q=0$, while sub-Poissonian processes are characterized by a negative Mandel parameter $Q<0$ (as, for example, in the case of the photon arrival statistics of squeezed light), with $Q=-1$ in the limit of complete suppression of fluctuations. When $Q>0$ one speaks of a super-Poissonian counting statistics, which can be the result of technical noise or of enhanced fluctuations intrinsic to the dynamics of the system (see discussion below). For detection efficiencies $\eta<100\%$ , the observed value of the Mandel parameter is $Q_{D}=\eta{}Q$ \cite{CubelLiebisch2005,Reinhard2008}. This detected value is used in all experimental figures in this letter. The Supplemental Material reports our experimental verification that a Poissonian distribution $Q_D=0$ is measured in the direct photoionization process, in which case no suppression of fluctuations is expected.\\
\begin{figure}[htp]
\includegraphics[width=0.5\linewidth]{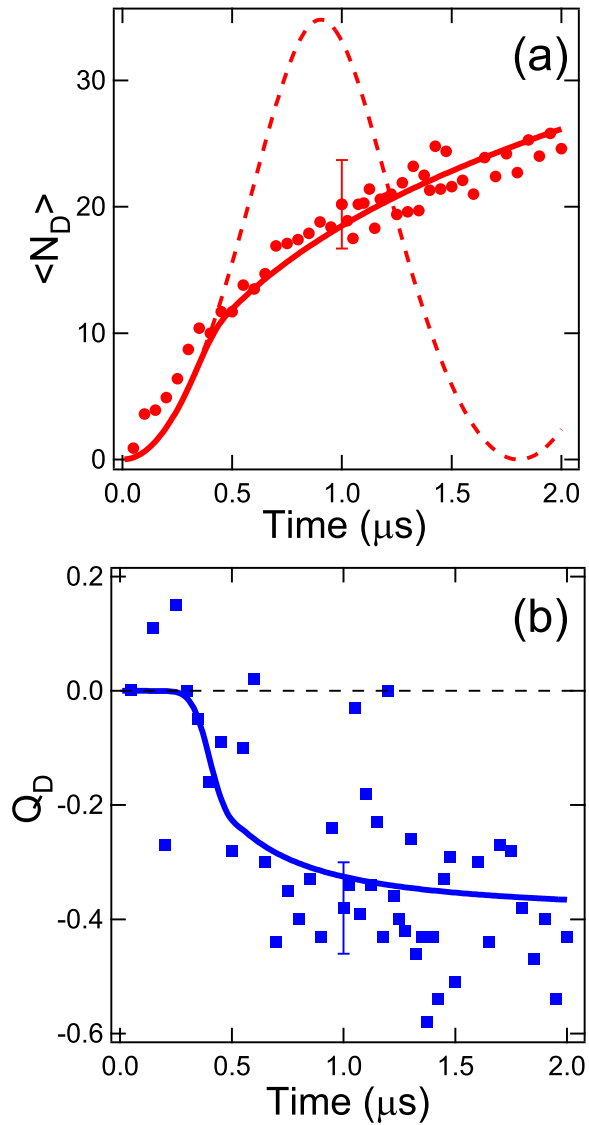}
\caption{Rydberg excitation dynamics in an ultra-cold gas. The mean number of detected ions $\langle N_D\rangle $ and $Q_D$ are shown, in a and b, respectively,  as a function of pulse duration for the resonant excitation of the  $71\mathrm{D}_{5/2}$ state. The blue laser detuning from the intermediate 6P$_{3/2}$ F=3 state is 1 GHz. The number of ground state atoms is $8\times{}10^3$, the average density $1.2\times10^{10}$ cm$^{-3}$ and the two-photon Rabi frequency is 40 kHz. The error bars are derived from 50  experimental realizations. The continuous lines are the prediction of our complete model, while the dashed line in a is the theoretical prediction of a model based on the Dicke symmetric states only. All quantities are rescaled by the detection efficiency.}
\label{Fig_Pulse}
\end{figure}
\indent
In Fig. \ref{Fig_Pulse}, an example of the excitation dynamics to the $71\mathrm{D}_{5/2}$ state for atoms on resonance is shown. Initially, a rapid increase in the average detected ion number is concomitant with a decrease in $Q_D$. After 0.5~$\mu$s the growth in the detected ion number slows down while $Q_D$ fluctuates around $-0.4$. Taking into account our detector efficiency of $\approx 40\%$~\cite{Viteau2010}, the measurements indicate that on resonance a value close to $Q=-1$ is obtained.
In the long time limit the effect of a varying Rabi frequency across the sample, due to both the distribution of atoms and the spatial laser profile and line-width, cause a dephasing of the excitations, leading to a steady growth in the number of Rydberg atoms excited. In order to obtain reliable statistics with a resonable number of experimental realizations, we performed numerical simulations to verify how quickly the mean and standard deviation converge for Poissonian and sub-Poissonian distributions. For example, to measure the Mandel parameter of a sub-Poissonian process with $Q\approx-0.5$ to within $\pm 0.1$ requires around $50$ realizations. \\
\begin{figure}[htp]
\includegraphics[width=0.5\linewidth]{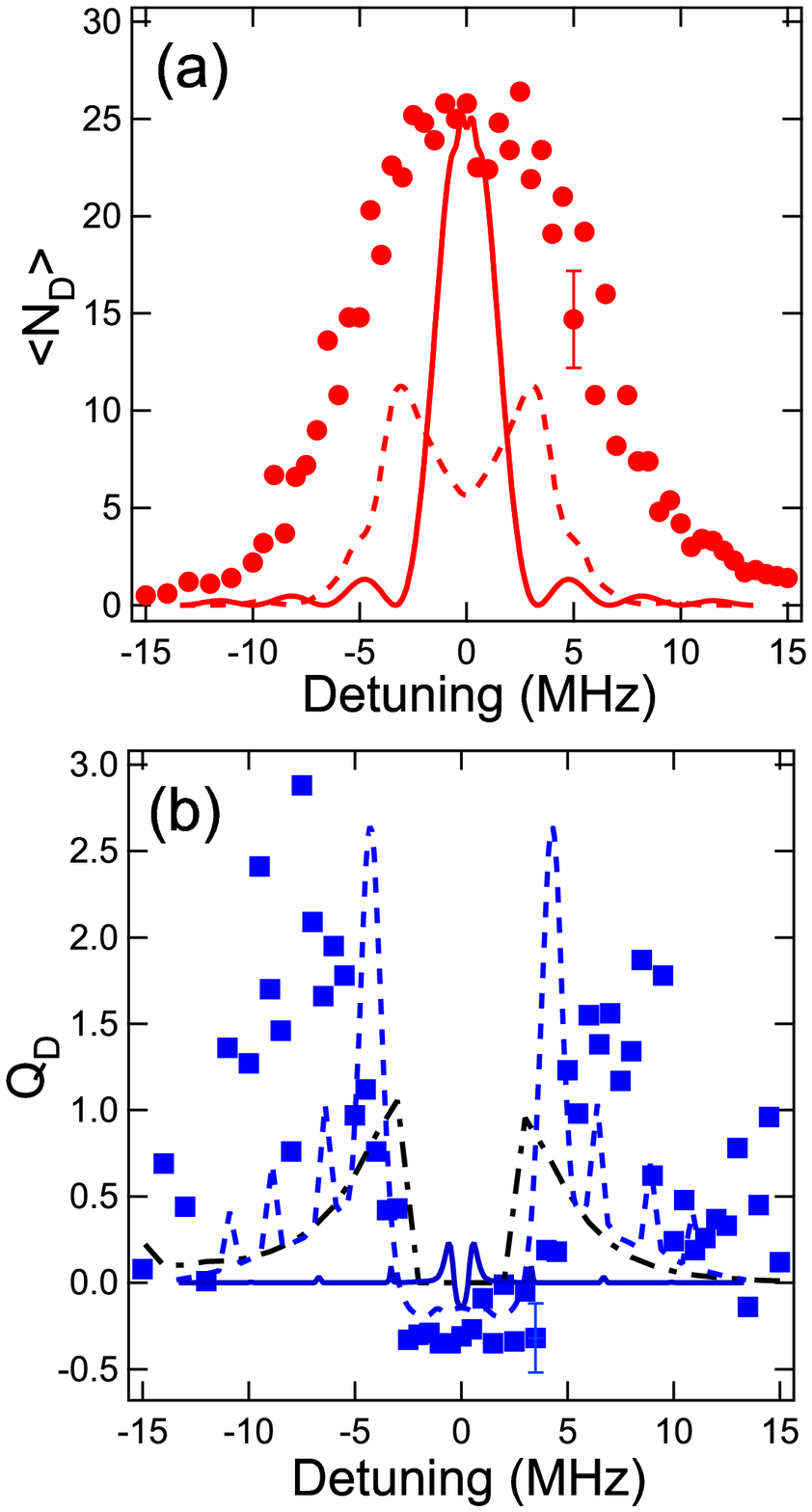}
\caption{Detuning-dependence of collective Rydberg excitations. The mean number of detected ions $\langle N_D\rangle $ and $Q_D$ are shown for a range of frequencies around the two-photon excitation of state $71\mathrm{D}_{1/2}$ in a and b, respectively.
The $\langle N_D\rangle$ data are well fitted by a top truncated Lorentzian lineshape, except in a frequency range around the maximum.
The continuous and dashed lines are the predictions of our complete model, with the calculated average coupling strength and 100 times that coupling strength, respectively.  The black dot-dashed line simulates the effect of the laser jitter (assumed to be $0.5\,\mathrm{MHz}$) on $Q_D$, assuming that $Q=0$ independently of the detuning, based on the data of Fig. 2a. The blue laser detuning from the intermediate 6P$_{3/2}$ F=3 state is 1 GHz. The number of ground state atoms is $5\times{}10^4$, the average density $4\times10^{10}$cm$^{-3}$, the two-photon Rabi frequency 40 kHz and the excitation pulse duration 0.3 $\mu$s.}
\label{Fig_Scan}.
\end{figure}
\indent The resonant excitation dynamics can be easily visualized as a saturation of the maximum number of superatoms that fit into the volume of the MOT, leading to a limiting value for $Q$ around $-0.7$ in a `close-packing' geometry of blockade spheres.
Away from resonance, however, the picture is more complicated. Figure~\ref{Fig_Scan} shows the mean of the detected ion number and of $Q_D$ as a function of the detuning from resonance for a fixed time. As expected, for zero detuning $Q_D$ is highly negative, i.e. close to $-1$. Away from resonance, however, the counting statistics quickly becomes super-Poissonian with $Q_D$ as large as $2-3$. One reason for this may be the effect of laser frequency fluctuations on the time scale of
the data acquisition periods (around 30~seconds). On either slope a small shift in frequency corresponds to a large difference in the number of detected ions, leading to a large variance. Our theoretical model discussed below, however, indicates that the measured positive value of the Mandel parameter off resonance is also intrinsic to the excitation dynamics of the system.  \\
\indent For a better understanding of the physics underlying our experimental observations we need a theoretical treatment of our system that gives access to the full counting statistics, and hence to the $Q$-parameter, both on and off resonance. Although recent theoretical work has addressed the expected counting statistics for on-resonant excitation of collective Rydberg states \cite{Ates2006,Wuester2010}, so far there have been no predictions for the off-resonant case. We describe our system as an ensemble of $N_0$ indistinguishable atoms, for which the laser excitation results in a single wavefunction superposition of a two level system composed of the ground state $|g\rangle$  and the Rydberg state $|R\rangle$, driven by a laser with Rabi frequency $\Omega$ and detuning $\delta$. Our approach is based on  the  DCS~\cite{Dicke1954,MandelWolf1995}, characterized by the cooperative number $r$ defining the symmetry and the  number $N$ of Rydberg excited atoms, $| r,N\rangle$.  As the laser excitation preserves the DCS symmetry, only states with the same symmetry number $r$ are excited when starting from an ultra-cold atomic sample described by a totally symmetric DCS. When vdW interactions are taken into account, states with the same symmetry and those with different symmetries can be coupled. \\
\indent  Assuming  for the initial atomic wave-function the fully symmetrical
 non-degenerate DCS $\left\vert
r=N_0/2,N\right\rangle =\left\vert N\right\rangle_s$ and neglecting the vdW coupling between symmetrical and antisymmetrical DCS, we write the atomic wavefunction as $\left\vert \Psi \left( t\right) \right\rangle =\sum_{N=0}^{N_0} a_{N}\left( t\right) \left\vert N\right\rangle_s.$
The time evolution of the amplitudes $a_N$  is given by the following coupled equations \cite{haroche2006} (with $\hbar=1$): %
\begin{eqnarray}
&{\dot a_{N}}=iN\delta a_{N} \\ \nonumber
&-i\frac{\Omega  }{2}\left[\sqrt{\left( N_0-N\right) \left( N+1\right) }%
a_{N+1}+\sqrt{N\left( N_0-N+1\right)}a_{N-1}\right].
\label{eqsymmetric}
\end{eqnarray}
The WdV interaction $W^N_{ss}=W_0N\left(N -1\right)/(N_0
(N_0-1))$ introduces an additional term
\begin{equation}
-iW^N_{ss}a_{N}
\label{WdVsymmetric}
\end{equation}
to the equation of motion for ${a_N}$,
where $W_0$ is the sum over all the interaction energies between pairs of Rydberg atoms~\cite{Tong2004}.
\indent The number of excitations in the Dicke ladder is equal to the number of Rydberg excitations, that is, a few tens in our experiment. The solution of the above equations allows one to calculate the cooperative mean and variance values, and all higher moments of the number of Rydberg excitations.  It  leads to the Rydberg excitation depicted by the dashed line of Fig. \ref{Fig_Pulse}, in reasonable agreement with the
experimental observations for short times, but not in the long time limit. In addition it predicts asymmetric profiles for the laser detuning dependence in the blockade regime,
not matching previous observations~\cite{Amthor2007} and our data on the detected ions vs the laser detuning shown in Fig.~\ref{Fig_Scan}a.\\
\indent The problem is that the vdW interaction couples the fully symmetrical DCSs to the remaining
DCSs within a fixed $N$ subspace, which has
a large degeneracy given by the binomial coefficient $C_{N_0}^{N}$. The most appropriate orthonormal basis within that subspace is composed of the symmetrical DCSs $\ \left\vert N\right\rangle_s$  and the superposition of non-symmetrical states $\left\{ \left\vert N,q \right\rangle_{ns} \right\}$, with $1\le q \le C_{N_0}^{N}-1$, for which  the vdW interaction matrix is diagonal. This original basis construction, detailed in the Supplemental Material, is far more general than the treatment of the vdW interactions provided we know
the basis diagonalizing the coupling. Within this basis the vdW matrix contains the terms $W^N_{ss},W^N_{qq},W^N_{sq}$, see Supplemental Material.\\
\indent By writing the enlarged atomic wave-function as a superposition of symmetric and non-symmetric states, we obtain that  the $a_N$ time evolution of
 Eq. (\ref{eqsymmetric}) is completed by terms describing the vdW coupling to the amplitude $b^q_N$  of the non-symmetrical state. As a first approximation, we neglect the laser excitation of the amplitudes $b^q_N$ because of their large degeneracy and the resulting
weak occupation for each of them. After the elimination of the equations for $b^q_N$, the temporal evolution of the symmetric states determined by the coupling with a bath, made up of the the remaining collective states, therefore contains both Markovian and non-Markovian memory function contributions. For our experimental conditions, this quite general result can be simplified when calculating the memory functions see Supplementary Material. In the end for each number $N$ of Rydberg excitations the  evolution  is described by a system of two equations, one for $a_N$ and one for a single level $\ c_N$ defined as the superposition of all asymmetric DCS's of  the fixed $N$ subspace.   Thus the bath is modeled by a number of states equal to that of the symmetric DCS's and the total required basis set required for the simulations  is approximately twice the number of Rydberg excitations.  The bath is Markovian and produces a shift of the symmetric DCS energy precisely compensating  the vdW eigenenergy of Eq.~\eqref{WdVsymmetric}. This compensation leads to symmetric excitation profiles as in Fig.~\ref{Fig_Scan}.  The vdW interaction for the
symmetric DCS finally replaces Eq.~\eqref{WdVsymmetric} by the following one:%
\begin{equation}
{\dot a}_{N} =-iW^N_{ss}c_{N}, \\
\end{equation}
For the non-symmetrical DCS
\begin{equation}
{\dot c}_{N} =i\delta Nc_{N}-iW^N_{ss}a_{N}.
\label{system}
\end{equation}
Notice that within the above equations the vdW interaction acts as a coherent coupling between symmetrical and antisymmetrical states, leading to a variety of coherent evolution features, for instance the revivals appearing as narrow peaks in the $Q$ dependence on parameters as the laser detuning, see Fig.~\ref{Fig_Scan}, or the laser pulse duration. However for each $N$ manifold the revival conditions are slightly different, and for a large number of Rydberg excitations, the revivals are washed out.  The last step of this theoretical analysis is the introduction of the laser
excitation into evolution of $c_N$.
We treat it perturbatively, with a limitation on the temporal evolution imposed by the  spectral linewidth $\Delta \omega$ of the excitation laser.  The final equations are reported within the Supplemental Material.\\
\begin{figure}[htp]
\includegraphics[width=0.5\linewidth]{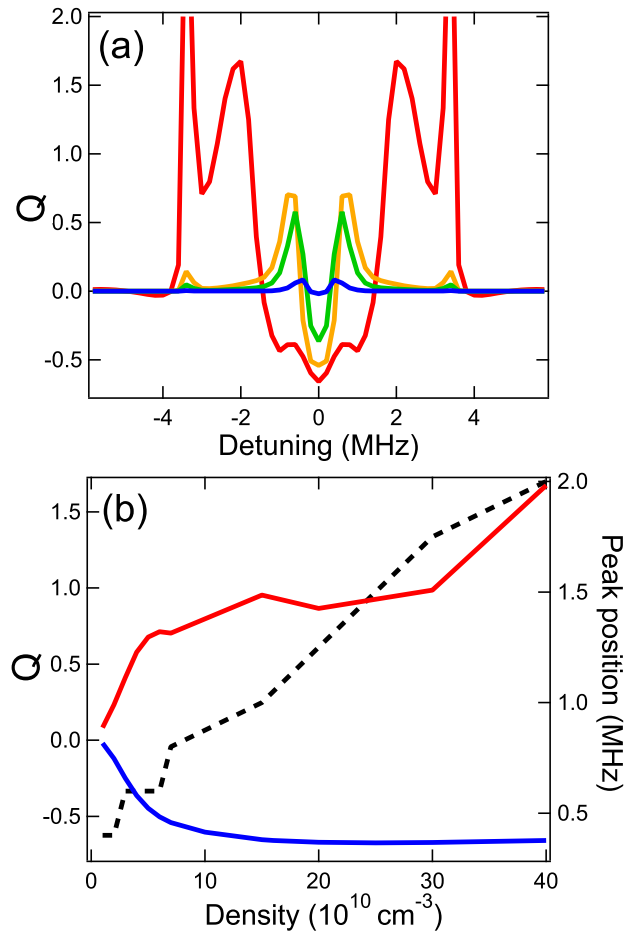}
\caption{Predictions for the detuning-dependence of collective Rydberg excitations. (a) Theoretical dependence of the $Q$-parameter on the laser detuning for a range of atomic densities (in cm$^{-3}$):   $1\times10^{10}$  (blue), $4\times10^{10}$ (green), $10\times10^{10}$ (orange) and $40\times10^{10}$  (red), respectively . (b) Dependence on the density of the on-resonant $Q$  (blue line) and the value of $Q$ (red line) and position (black dashed line, right-hand scale) of the first off-resonant peaks. }
\label{Fig_Q}
\end{figure}
\indent Typical results of numerical simulations using the Dicke model are shown in Figs. \ref{Fig_Pulse} and \ref{Fig_Scan} together with our experimental data. Clearly, the salient features of the experiment are reproduced well by our model, both in the dynamics  and in the dependence on the detuning. In particular, the off-resonant behaviour of the $Q$-factor, which in the experimental data features two symmetric peaks, is also visible in the simulation.  The variation in the interatomic interaction curves, see Supplemental Material, is reflected in the uncertainty in our determination of the coupling strength for the theoretical curves in Fig.~\ref{Fig_Scan}. Perfect quantitative agreement between experiment and theory is not expected as the Dicke model does not, at this time, include the spatially varying atomic density of the MOT, Doppler broadening, the spatial inhomogeneity of the excitation lasers, the full manifold of sublevels of the Rydberg states, and the effects of magnetic and possible residual electric fields, all of which might contribute to the larger linewidth observed experimentally. In particular, the dip in the Rydberg number around zero detuning seen in Fig.\ref{Fig_Scan} (a) could be smoothed out when taking into account the density variation over the cloud. In fact, Figure 3 shows that the variation of $Q$ with the detuning depends strongly on the density (or, equivalently, on the $C_6$ coefficient that describes the van-der-Waals interaction between Rydberg atoms). As the density is increased, the value of $Q$ on resonance goes from zero to a limiting value of around $-0.7$, whereas the position and height of the off-resonant peaks symmetric around zero detuning increase steadily. In order to get better agreement, one could either perform experiments with close to uniform atomic densities (e.g., using flat-top excitation beams much smaller than the atomic sample) or develop a theoretical description that allows for the non-uniform density to be included. \\
\indent In summary, we have shown clear evidence of sub-Poissonian counting statistics of Rydberg excitations in an ultra-cold atomic sample, which are a clear signature of many-body quantum correlations in such a system. We have characterized both the dynamics and the dependence on the detuning of the $Q$-parameter and reproduced the main features of both using a novel model. In future studies, it will be interesting to look directly at the spatial correlations between Rydberg excitations~\cite{Schwarzkopf2011}, at the counting statistics for the antiblockade regime of~\cite{Amthor2010}, and at other, more indirect, signatures of Rydberg-Rydberg interactions, for example in EIT experiments~\cite{Pritchard2010}.\\
\indent This work was supported by the E.U. through grants No. 225187-NAMEQUAM and No. 265031-ITN-COHERENCE and the collaboration between University of Pisa and University of Paris Sud-11. The authors thank M. Allegrini, R. C\^{o}t\'{e},  P. Grangier, T. Pohl, M. Saffman, J. Stanojevic and M. Weidem\"uller for useful discussions and J. Radogostowicz for assistance.

\section*{Supplemental Material}
Sec.~I provides additional information on the experimental setup, on the reliability of the measurements yielding negative values for $Q_D$ and on the absence of ions within the laser excitation volume. Within the theoretical description based on the DCS, after a brief recall of the mean-field Sec.~II describes the construction of a basis of
non-symmetric states the construction of the bath of the non-symmetric states and their excitation by the laser. Sec.~III discusses the vdW parameters determining the Rydberg interactions.

\section{Set-up}
\indent  Experiments are performed using a two-photon excitation scheme for Rydberg excitations in a small cloud of around $10^5$ $^{87}$Rb atoms trapped in a MOT of radius around 30~$\mu{}$m and densities between $1\times10^{10}$ and $5\times10^{10}$ atoms/cm$^{3}$ determined to within a factor of $2$ from fluorescence images (for details see \cite{Viteau2010}). The sizes of the excitation beams (around 200~$\mu{}$m) are chosen such as to obtain intense beams that are approximately uniform across the atomic sample. Single atom two-photon Rabi frequencies of around 100 kHz are achieved. The excitation lasers are pulsed for up to $20\,\mathrm{\mu s}$, after which the Rydberg atoms are field ionized and the ions are detected by a channeltron. For the two-photon excitation, a laser beam at 421~nm is detuned around 0.5--1~GHz from the $\mathrm{5S_{1/2}(F=2)}\rightarrow\mathrm{6P_{3/2}(F'=3)}$ transition, and an infra-red beam of wavelength 1013--1015~nm provides the second step to the Rybderg state $\mathrm{6P_{3/2}(F'=3)}\rightarrow\mathrm{nS/nD}$. Both of these beams are controlled using acousto-optic modulators (AOMs) with a rise-time of around 80~ns. Both lasers are stabilised to a scanning Fabry-Perot cavity to prevent long-term frequency drift. We estimate that the two-photon linewidth on the timescale of the excitation pulses is around 300~kHz. In order to achieve reliable statistics, the excitation sequence is repeated up to a few hundred times with the repetition rate limited to $10$ Hz in order to avoid a buildup of charge on the quartz cell surrounded by the external field plates \cite{Viteau2011}. The complete acquisition procedure takes around 30~s.
The detection efficiency for ions in our experiment is around $40\%$.\\
\indent In order to verify that the observed sub-Poissonian statistics for resonant excitation is actually due to collective Rydberg excitations and not an artefact of the limitations of the experimental apparatus, we performed a series of additional experiments. Most importantly, we performed an experiment that interpolated between Poissonian and sub-Poissonian statistics for a fixed average number of Rydberg excitations, thus ruling out a spurious effect due to a saturation of the channeltron. To that end, the volume of the MOT was varied using a combination of magnetic field gradients, trapping beam sizes and laser powers chosen such as to keep the number of excited Rydberg atoms constant. As expected, for a small cloud the Rydberg dynamics was in the highly collective regime and a negative $Q_D$ was measured. As the volume increased with the number of Rydberg excitations kept fixed, the excitation dynamics tended to a Poissonian distribution with $Q\approx 0$. Moreover, we performed experiments in which the first step laser was tuned to resonance with the $\mathrm{5S_{1/2}(F=2)}\rightarrow\mathrm{6P_{3/2}(F'=3)}$ transition so that atoms were directly ionized through the absorption of two photons at 421 nm. Up until counting saturation, this process was found to obey Poissonian statistics with $Q_{D}=0$.\\
\indent The creation of spurious ions, that is, those not created by field ionization of Rydberg atoms, is a delicate issue in studies of the dipole blockade. The Stark shift due to a single ion is larger than the van-der-Waals interaction between two atoms in the investigated $D_{5/2}$ state. Thus the creation of one ion in the sample can lead to a blockade due to Coulomb interactions which suppresses the excitation of other Rydberg atoms and thus mimicks the dipole blockade. A possible cause for the presence of spurious ions is the absorption of two 421~nm photons leading to direct ionization from the intermediate 6$^2$P$_{3/2}$ state \cite{Viteau2011}. By detuning the blue laser 1~GHz from this state excitation the probability of creating an ion is $3\times 10^{-7}$ per $2\,\mathrm{\mu s}$ pulse. For the low atoms numbers in our experiment this equates to around $3\times 10^{-2}$ per shot and thus is not a problem. Also, black-body radiation limits the lifetime of Rydberg atoms, for the 71$D_{5/2}$ state the effective lifetime is $\sim$150$~\mu{}$s~\cite{Beterov2009}; however, the excitation and ionization is performed within $<10~\mu$s and thus should have negligible impact on the counting statistics. Finally, at the typical atomic densities used in the experiment the decay process is not affected by super-radiance. Any loss from the initially prepared state will, therefore, follow Poissonian statistics.

\section{Theory}
\label{theory}
\subsection{Mean field}
In a many-body system with interactions, the mean-field theory replaces all interatomic interactions with an average or effective interaction, reducing the many-body problem to an effective one-body problem. Within this approach the Rydberg state energy is shifted by the vdW interaction characterized by the parameter $W_{0}$ given by a sum over all the  Rydberg pairs inside the atomic cloud, see \cite{Tong2004}.  For instance, in a medium with uniform atomic density $n$ and for a $C_6$ interaction depending as $r^{-6}$ on the interatomic distance $r$,  $W_0$ is approximatively  $\frac{C_{6}}{<r>^6}\approx C_{6}\left( \frac{4\pi }{3}\right)^{2}n^2$.

\subsection{vdW basis in the cooperative approach}
The vdW interaction couples the fully symmetrical DCSs to the remaining
DCSs within the fixed $N$ subspace  having
the large $C_{N_0}^{N}$ degeneracy. The most appropriate orthonormal basis within that subspace is composed of the symmetrical DCSs $ \left\vert N\right\rangle_s$  and the superposition of non-symmetrical states $\left\{ \left\vert N,q \right\rangle_{ns} \right\}$, with $1\le q \le C_{N_0}^{N}-1$, for which  the vdW interaction matrix is diagonal.   To obtain the correct number of states within the $C_{N_0}^{N}$ ensemble, we first remove from the $\left\{ \left\vert N,q\right\rangle \right\}$ set one state labelled  $\left\vert N,p\right\rangle $, assuming $W_{pp}^{N}=W_{ss}^{N}$. Considering
the state
\begin{equation}
\left\vert N,{\widetilde q}\right\rangle =\left\vert N,q%
\right\rangle +\frac{1}{\sqrt{C_{N_0}^{N}}-1} \left\vert N,p\right\rangle,
\end{equation}
for  $1\le q \le C_{N_0}^{N}-1$. Then the basis is constructed by subtracting for each $\vert N,{\widetilde q}\rangle $ state  its projection on the symmetrical
$\left\vert N\right\rangle _{s}$ state. We obtain the $\left\{ \left\vert N,q \right\rangle_{ns} \right\} $  states
defined as%
\begin{equation}
\left\vert N,q\right\rangle _{ns}=\left[ \left\vert N,{\widetilde q}
\right\rangle -_{s}\left\langle N\mid N,{\widetilde q}\right\rangle \left\vert
N\right\rangle _{s}\right] .
\end{equation}
It is easy to verify that such a basis $\left\{ \left\vert N\right\rangle
_{s},\left\{ \left\vert N,q\right\rangle _{ns}\right\} \right\} $ is
orthonormal. The van der Waals coupling is diagonal with eigenvalue $W_{qq}^{N}$ for the restricted basis $\left\{ \left\vert N,q\right\rangle _{ns}\right\} $; these eigenvalues are not modified by the above basis changes. The coupling between the states $\left\vert N\right\rangle
_{s}$\ and $\left\vert N,q\right\rangle _{ns}$ is%
\begin{equation}
W_{sq}^{N}=_{s}\left\langle N\right\vert W\left\vert N,q\right\rangle _{ns}=%
\frac{1}{\sqrt{C_{N}^{N}}}\left[ W_{qq}^{N}-W_{ss}^{N}\right] .
\label{Wenergies}
\end{equation}
\indent The above construction of the basis is very general, provided we know
the basis which diagonalizes the coupling. The only approximation is the
existence of the state  $\left\vert N,p\right\rangle $. For
a large number of atoms, such an approximation is reasonable because it is
possible to identify a state $\left\vert N,\widetilde{p}\right\rangle $ such that $W_{{\widetilde p}{\widetilde p}}^{N}\approx W_{ss}^{N}$.

\subsection{Enlarged wavefunction evolution}
By writing the enlarged atomic wave-function as
\begin{equation}
\left\vert \Psi \left( t \right) \right\rangle =\sum_{N=0}^{N_0} a_{N}\left( t\right) \left\vert N\right\rangle_s + \sum_{q} b^q_{N }\left( t\right) \left\vert
N,q \right\rangle_{ns},
\end{equation}  the $a_N$ time evolution of
 Eqs. (2) and (3) in the main text  is completed by terms describing the  ($b^q_N, a_N$) vdW coupling. As a first approximation, we neglect the laser excitation of the amplitudes $b^q_N$ because of their large degeneracy and the resulting
weak occupation for each of them. Thus eliminating the equations for $b^q_N$, we obtain the following integro-differential equations:
\begin{eqnarray}
{\dot a}_{N} &=&-i(W_{ss}^N-N\delta)a_{N}-i\frac{\Omega  }{2}\left[\sqrt{\left( N_0-N\right) \left(
N+1\right) }a_{N+1}+\sqrt{N\left( N_0-N+1\right)}a_{N-1}\right] \nonumber \\
&+&\sum_q\int_{0}^{t} \left[ -\frac{d^{2}%
}{d\tau ^{2}}-2i\frac{d}{d\tau }W_{ss}^N+\left(W_{ss}^N\right)^{2}\right]   \frac{e^ { i\left(W_{qq}^N-N\delta\right)\tau}}{\sqrt{C_{N_0}^N}} a_{N}\left(
t-\tau \right) d\tau.
\label{differointegral}
\end{eqnarray}

\begin{figure}[htp]
\includegraphics[width=0.75\linewidth]{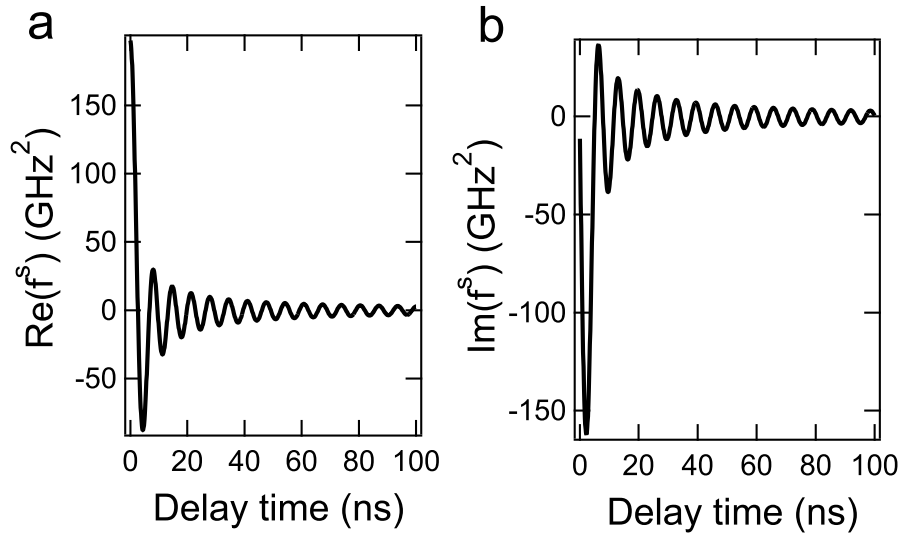}
\caption{Real and imaginary part of $f^s_{N}\left( t \right) $ vs the delay time $\tau$ for typical parameters investigated in our experiment.  }
\label{correlationfunction}
\end{figure}

The coupling $b_N \to a_N$ appearing in Eq. \eqref{differointegral} is described through the correlation function $f_N(\tau)$  as \begin{equation}
{\dot a_{N}}=-\int_{0}^{t}f_N(\tau) e^{iN\delta \tau} a_N\left(t-\tau \right) d\tau,
\end{equation} with  the correlation function given by
\begin{equation}
f_N\left( \tau \right) =\sum_q\frac{1}{C^N_{N_0}}\left[ \left[ -%
\frac{d^{2}}{d\tau ^{2}}-2i\frac{d}{d\tau }W^N_{ss}+(W^N_{ss})^{2}\right]e^
{ -iW^N_{qq}\tau} \right],
\end{equation}%
where $W^N_{sq}$ was eliminated using Eq. \eqref{Wenergies}. With only two Rydberg excited atoms  within the volume $V$, $f_N$ becomes
\begin{equation}
f_N\left( \tau \right) =\frac{W_0^{2}}{4V}\int_{2V/N}^{V}\left[
\frac{1}{v^{4}}-\frac{N}{v^{2}V^{2}}+\frac{N^{2}}{4V^{4}}\right]
e^{ -i\frac{2W_0}{v^{2}}\tau} dv.
\end{equation}
The above integral  has an analytical solution. The solution for the correlation function is composed of a long-memory component
$f^l_{N}\left( \tau \right) $ and a short-memory one, $f^s_{N}\left( \tau \right) $. The real and imaginary parts of $f^s_{N}\left( \tau \right) $ are represented in  Fig.~\ref{correlationfunction}. The $f^s_{N}\left( \tau \right) $ contribution treated in the Markov approximation leads to the following imaginary term:
\begin{equation}
-\int_{0}^{t}f^s_{N}\left( \tau \right) \exp{\left(i\delta j \tau\right)}a_{N}\left( t-\tau \right) d\tau
=iW^N_{ss}a_{N}\left( t\right).
\end{equation}%
This term exactly compensates exactly the blockade energy term introduced in Eq.~\eqref{differointegral} above for the symmetric DCS theory and, as a consequence, produces the symmetrical resonance excitation profiles reported in the Fig. 1 of the main text.\\
\indent The long-memory part can be treated exactly, but a good approximation, up to a few tens of Rydberg excitations, is
\begin{equation}
f^l_{N}\left( \tau \right) =(W^N_{ss})^{2}.
\end{equation}
Using the above short-memory and long-memory parts,
the vdW coupling can be described by representing the
ensemble of the non-symmetrical levels with $j$ Rydberg excitation in terms of a single
level $\ c_j$ defined as the superposition all asymmetric DCS's of  the fixed $N$ subspace. Within this simplified scheme for each number $N$ of Rydberg excitations the  evolution  is described by the following system of two equations, one for $a_N$ and one for a single level $\ c_N$, reported also in the text:
\begin{eqnarray}
{\dot a_{N}}&=&-iN\delta  a_{N}
-i\frac{\Omega  }{2}\left[\sqrt{\left( N_0-N\right) \left( N+1\right)}
a_{j+1}+\sqrt{N\left( N_0-N+1\right)}a_{N-1}\right] -iW_{ss}^Nc_{N}, \label{equationA} \\
{\dot c}_{N} &=&-iN\delta c_{N}-iW_{ss}^Na_{N}.
\label{equationC}
\end{eqnarray}
\subsection{Laser excitation of the non-symmetric bath}
The last step of this theoretical analysis is the introduction of the laser
excitation into evolution of $c_N$.
We treat it perturbatively, with a limitation on the temporal evolution imposed by the  spectral linewidth $\Delta \omega$ of the excitation laser.  The final result is a modification of Eq. \eqref{equationC} into \begin{equation}
{\dot c}_{N} =-\left(\frac{\Gamma_N}{2}+iN\delta\right) c_{N}-iW_{ss}^Na_{N} +\frac{\Gamma_N}{2} \left|c_{N-1}\right|,
\label{FinalEquation}
\end{equation}
with all coherence terms $c_Nc^*_{N-1}$ set to zero.  $\Gamma_N$ depends on the laser linewidth $\Delta \omega$ as
\begin{equation}
\Gamma_N=\frac{\Omega^2(N_0-2N)}{4}\left[\frac{\Delta \omega/2}{\left(\Delta \omega/2\right)^2+\left(\delta-NW_{ss}\right)^2}+\frac{\Delta \omega/2}{\left(\Delta \omega/2\right)^2+\left(\delta+NW_{ss}\right)^2}\right].
\end{equation}

\section{Rydberg interaction parameters}
\label{parameters}
The numerical results rely on the precise values for the vdW coupling strengths. In a first approximation, we can consider that, in zero field, two Rydberg atoms  experience a ${C_6}/{r^6}$ interaction, $C_6$ being  calculated on the basis of  perturbation theory of \cite{SInger2005}. This model fails to describe the interaction accurately, first at small interatomic distances and secondly because of the Zeeman degeneracy of the two-atom Rydberg state. While at large interatomic distances, the two Rydberg states are dipole coupled via one additional state, at very short interatomic distances a large number of states is coupled, and the overall coupling is reduced. In our calculation of the interaction energy $W_0$, we introduce an appropriate cut-off radius (roughly twice the atomic size) because the excitation of two Rydberg atoms having a distance smaller than the cut-off radius is very unlikely. At intermediate interatomic distances  the perturbative treatment of \cite{SInger2005} is not fully valid because of crossing in the molecular levels. Then a full diagonalization of the Hamiltonian leads to a more complex dependence of the interatomic coupling on the interatomic distance,  ${C_6}/{r^6}$ at large distance and ${C_3}/{r^3}$, i.e. a dipole-dipole interaction, at smaller ones. Finally, Walker and Saffman~\cite{WalkerSaffman2008} have shown the Rydberg interaction is strongly dependent of the relative orientation of magnetic momentum of the colliding atoms of the pair. For the of $D_{\frac{5}{2}}$ states this leads to a variety of 21 possible interaction curves whose strengths vary by more than two orders of magnitude. The average presented in~\cite{WalkerSaffman2008} allowed us to obtain an effective dipole-dipole interaction, i.e., a single interaction curve for the  calculation of the parameter $W_0$. The variation in the interaction curves is reflected in the uncertainty in our determination of the coupling strength for the theoretical curves in Fig. 1 of the main text.


\begin{thebibliography}{99}
\bibitem{Saffman2010} M. Saffman, T.G. Walker, and K. M{\o}lmer, Rev. Mod. Phys. {\bf 82}, 2313 (2010).

\bibitem{Comparat2010} D. Comparat and P. Pillet,  J. Opt. Soc. Am. B {\bf 6}, A 208 (2010).

\bibitem{LesanovskyViewpoint2011} I. Lesanovsky, Physics {\bf 4}, 71 (2011).

\bibitem{CubelLiebisch2005} T. Cubel Liebisch, A. Reinhard, P.R. Berman, and G. Raithel,  {\em Phys. Rev. Lett.} {\bf 95}, 253002 (2005) and Erratum {\em  ibid.} {\bf 98}, 109903 (2007).

\bibitem{Reinhard2008} A. Reinhard, K.C. Younge, and G. Raithel, Phys. Rev. A {\bf 78}, 060702(R) (2008).

\bibitem{Gaetan2009} A. Ga\"etan {\em et al.},  Nature Phys. {\bf 5}, 115 (2009).

\bibitem{Urban2009} E. Urban {\em et al.},  Nature Phys. {\bf 5}, 110 (2009).

\bibitem{Hernandez2008} J.V. Hern\'andez, and F. Robicheaux, J. Phys. B: At. Mol. Opt. Phys. {\bf 41}, 5301(2008).

\bibitem{Tong2004} D. Tong {\em et al.}, Phys. Rev. Lett. {\bf 93}, 063001 (2004).

\bibitem{Ates2006} C. Ates, T. Pohl, T. Pattard, and J.M. Rost, J. Phys. B: At. Mol. Opt. Phys. {\bf 39}, L233 (2006).

\bibitem{Wuester2010} S. W\"{u}ster {\em et al.}, Phys. Rev. A {\bf 81} 023406 (2010).

\bibitem{Stanojevic2010} J. Stanojevic and R. C\^{o}t\'e, Phys. Rev. A  {\bf 81}, 053406(2010).

\bibitem{Lesanovsky2011} I. Lesanovsky, Phys. Rev. Lett. {\bf 106}, 025301 (2011).

\bibitem{Hernandez2006} J.V. Hernandez and F. Robicheaux, J. Phys. B {\bf 39}, 4883 (2006).

\bibitem{Heidemannn07} R. Heidemann {\em et al.},  Phys. Rev. Lett. {\bf 99}, 163601 (2007).

\bibitem{Schwarzkopf2011} A. Schwarzkopf, R.E. Sapiro, and G. Raithel, Phys. Rev. Lett. {\bf 107}, 103001 (2011).

\bibitem{Dicke1954} R.H. Dicke, Phys. Rev. {\bf 93}, 99 (1954).

\bibitem{Pavolini1985} Pavolini, D., Crubellier, A., Pillet, P., Cabaret, L. and Liberman, S. {\em Phys. Rev. Lett.} {\bf 54}, 1917 (1985).


\bibitem{Viteau2010} M. Viteau {\em et al.}, J. Phys. B: At. Mol. Opt. Phys. {\bf 43}, 155301 (2010) and Erratum {\em  ibid.} {\bf 44} 079802 (2011).

\bibitem{Mandel1979} L. Mandel, Phys. Rev. Lett. {\bf 49}, 136 (1982).

\bibitem{MandelWolf1995}  L. Mandel and E. Wolf,  {\it Optical Coherence and Quantum Optics}, (Cambridge University, 1995).
\bibitem{haroche2006} S. Haroche and J.M. Raimond,{\it Exploring the quantum : atoms, cavities and photons}, p. 220 (Oxford, 2006).
\bibitem{Amthor2007} T. Amthor, M. Reetz-Lamour, S. Westermann, J. Denskat, and M. Weidem\"uller, Phys. Rev. Lett. {\bf  98}, 023004 (2007).
\bibitem{Amthor2010} T. Amthor, C. Giese, C.S. Hofmann, and M. Weidem\"uller, Phys. Rev. Lett. {\bf  104}, 013001 (2010).
\bibitem{Pritchard2010} D.J. Pritchard {\em et al.}, Phys. Rev. Lett. {\bf105}, 193603 (2010).

\bibitem{Viteau2011} Viteau, M. {\em et al.}, Opt. Express  {\bf 19}, 6007 (2011).

\bibitem{Beterov2009}  Beterov I.I., Ryabtsev I.I., Tretyakov D.B. and Entin, V.M., Phys. Rev. A {\bf 79}, 052504 (2009).

\bibitem{Tong2004} D. Tong {\em et al.}, Phys. Rev. Lett. {\bf 93}, 063001 (2004).

\bibitem{Itano1993} Itano, W.M. {\em et al.},  Phys. Rev. A  {\bf 47}, 3554 (1993).

\bibitem{Hernandez2008} Hern\'andez, J.V. \and Robicheaux, F., J. Phys. B: At. Mol. Opt. Phys. {\bf 41}, 5301(2008).



\bibitem{SInger2005} Singer, K. {\em et al.}, J. Phys. B: At. Mol. Opt. Phys. {\bf 38}, S321 (2005).

\bibitem{WalkerSaffman2008} Walker, T.G. and Saffman, M. Phys. Rev.  {\bf 77}, 032723 (2008).
\end{thebibliography}
\end{document}